# Estimating real-world probabilities:
# A forward-looking behavioral framework

Ricardo Crisóstomo[*]

26 November 2020

## Abstract

We show that disentangling sentiment-induced biases from fundamental expectations significantly improves the accuracy and consistency of probabilistic forecasts. Using data from 1994 to 2017, we analyze 15 stochastic models and risk-preference combinations and in all possible cases a simple behavioral transformation delivers substantial forecast gains. Our results are robust across different evaluation methods, risk-preference hypotheses and sentiment calibrations, demonstrating that behavioral effects can be effectively used to forecast asset prices. Further analyses confirm that our real-world densities outperform densities recalibrated to avoid past mistakes and improve predictive models where risk aversion is dynamically estimated from option prices.

[*] Corresponding author. Comisión Nacional del Mercado de Valores (CNMV), Edison 4, 28006 Madrid, Spain; and National Distance Education University (UNED), Bravo Murillo 38, Madrid, Spain. *e-mail*: rcayala@cnmv.es. The author acknowledges useful comments from Javier Ojea.

**List of contents**

## 1. Introduction

Asset pricing models have evolved under the paradigms of market efficiency and rational expectations. Yet, since the pioneering work of Keynes (1936), increasing evidence shows that investors commit systematic behavioral errors that manifest through asset prices[1].

We contribute to the literature by developing a forward-looking framework to measure investor sentiment and quantify its effects. Methodologically, we start with the risk-neutral distributions obtained from the most common benchmarks in financial economics, including stochastic volatility models, discontinuous jumps and non-parametric densities. All risk-neutral predictions are adjusted to incorporate investor's risk preferences through several utility formulations.

We next estimate the sentiment function which summarizes investor biases in specific areas of the return distribution. Following Cochrane (2005) and Shefrin (2008), we analyze the impact of behavioral biases through the Stochastic Discount Factor (SDF). The SDF or pricing kernel is the cornerstone of asset pricing, embodying investor preferences and beliefs about future returns. In traditional finance, the SDF must be monotonically decreasing, reflecting a diminishing marginal utility in terms of wealth. However, empirical analyses show that the SDF exhibits a counterintuitive upward-sloping portion, giving rise to the *pricing kernel puzzle* (see Aït-Sahalia and Lo, 2000; Jackwerth, 2000; and Rosenberg and Engle, 2002)[2].

When the SDF is expanded to incorporate sentiment effects, the pricing kernel collectively embodies time-discount, risk preferences and behavioral biases. While the first two are well-known in finance, the behavioral component of the SDF represents the change of measure required to incorporate investor sentiment in different areas of the probabilistic forecast. Following Barone-Adesi et al. (2017), we consider three possible sentiment-induced mistakes: Excessive optimism, which generates biases in average returns; overconfidence, which impacts volatility predictions, and tail sentiment, which is related to non-rational tail expectations.

We calibrate our sentiment function using forward-looking market-based inputs. Investor optimism is derived from changes in implied volatilities; overconfidence is proxied by changes in trading volumes, and tail sentiment is obtained from the skewness of the risk-neutral distribution. All investor biases are then aggregated into an *ex-ante* sentiment function that is used to transform risk-adjusted forecasts into real-world densities. To gauge the impact of sentiment effects, we consider two alternative calibrations: a low impact and a high impact specification.

We then examine the out-of-sample performance of all density forecasts. The accuracy of each model is assessed through the log-likelihood score; forecast errors are evaluated in terms of the Continuous Ranked Probability Score (CRPS), and statistical consistency is measured with the Berkowitz, Jarque-Bera and Kolmogorov-Smirnov tests. Finally, we summarize all forecast metrics in a standarized ranking using the Integrated Forecast Score (IFS).

---

[1] See, among others: De Long et al. (1990), Bollen and Whaley (2004), Constantinides et al. (2009), Baker and Wurgler (2007), Baker and Wurgler (2006), Han (2008), Yan (2010), Stambaugh et al. (2012), and Da et al. (2015).

[2] For a comprehensive survey of the pricing kernel puzzle, see Cuesdeanu and Jackwerth (2018).

Our results are striking. We analyze 15 stochastic models and risk-preference combinations and in all possible cases a simple behavioral correction generates substantial forecast gains. Remarkably, the improvement delivered by our real-world densities is robust across all evaluation methods, risk-preference hypotheses and sentiment calibrations, demonstrating that sentiment effects can be effectively used to forecast future prices.

We also perform two additional tests. First, we show that our real-world forecasts non-parametric densities that have been recalibrated to avoid past mistakes. Second, we show that behavioral corrections also improve the explanatory power of density predictions where risk aversion is dynamically estimated from option prices.

**Novelty and related research**

Compared to existing literature, the novelty of our approach is fourfold:

1. The traditional method to compute the SDF compares a backward-looking forecast with a forward-looking RND, questioning whether the pricing kernel puzzle could be caused by misaligned expectations[3]. While several papers address this information mismatch[4], we are the first to estimate the real-world pricing kernel, including both risk-preferences and behavioral effects, in a consistent forward-looking framework.

2. Previous studies attribute to sentiment all deviations between a traditional kernel and the empirical SDF. While straightforward, this approach does not consider whether such differences are caused by sentiment or other model misspecifications. Alternatively, we develop an estimation method which generates the behavioral SDF in terms of simple market-based inputs.

3. Sentiment effects are typically analyzed in a mean-variance framework. However, it is well known that both asset returns and investor's beliefs exhibit non-normal features. To comprehensively examine how sentiment manifests in real-world predictions, we consider the impact of investor biases across the entire forecast distribution, incorporating non-Gaussian features: (i) at the RND estimation level, (ii) in the behavioral SDF computation and (iii) in the density evaluation framework.

4. Though several researchers explore how to improve the forecasting ability of RNDs, the proposed adjustments are generally calibrated in-sample[5]. In contrast, our behavioral estimates are obtained strictly out-of-sample, respecting real-world conditions by using only the information known at each forecast date.

---

[3] See Brown and Jackwerth (2012), Beare and Schmidt (2016), Beare (2011), Carr and Wu (2003), Grith, Härdle and Krätschmer (2017), and Yatchew and Härdle (2006).
[4] See Linn et al. (2018), Cuesdeanu and Jackwerth (2018) and Sala et al. (2016).
[5] See Bliss and Panigirtzoglou (2004), Liu et al. (2007), Kang and Kim (2006) and Alonso et al. (2009).

## 2. Estimating real-world densities

We estimate real-world probabilities using a 3-step process. First, we start with the RNDs obtained from the cross-section of option prices. Next, risk-neutral distributions are adjusted to incorporate investors risk preferences through several utility formulations. Finally, we develop a forward-looking framework to measure investor sentiment and quantify its effects.

A growing literature shows that option-implied distributions, even without risk-adjustments, outperform historical forecasts in information content[6]. Building on this insight, we examine whether the predictive ability of RNDs can be further improved by considering two attributes of real-world investors that manifest in asset prices:

- **Risk preferences:** Empirical papers show that when RNDs are adjusted to incorporate investor risk preferences, their explanatory power increases[7].

- **Behavioral biases:** Option prices reflect tailored views about specific areas of the return distribution, hence being prone to attract sentimental trades. If we accept that options are affected by sentiment, it follows that option-implied forecasts should be appropriately adjusted to disentangle investor biases from fundamental expectations.

### 2.1 Risk-neutral densities

We start with the risk-neutral forecasts obtained from the most common benchmarks in finance. Our first density is obtained from the lognormal Black-Scholes-Merton (LN) model. Second, we consider the Heston (1993) stochastic volatility dynamics:

$$dF_t = \sqrt{V_t} F_t dW_{t,1} \tag{1}$$

$$dV_t = a(\overline{V} - V_t)dt + \eta \sqrt{V_t} dW_{t,2} \tag{2}$$

where $F_t$ is the forward price at date $t$, $V_t$ and $\overline{V}$ represent the instantaneous and long-term variance, respectively, $a$ is the variance mean-reversion speed, $\eta$ is the volatility of the variance process and $dW_{t,1}$ and $dW_{t,2}$ are two correlated Wiener processes. Next, we complement the Heston stochastic volatility in (2) with discontinuous jumps, obtaining the Bates (1996) model:

$$dF_t = \sqrt{V_t} F_t dW_t^{(1)} + J_t F_t dN_t - \lambda \mu_J F_t dt \tag{3}$$

where $N_t$ is a Poisson process with intensity $\lambda$ and $J_t$ are lognormal jumps with mean $\mu_J$ and standard deviation $v_J$. Fourth, we consider the Variance Gamma (VG) process (Madan, Carr, and Chang, 1998), which combines small and large price jumps through the dynamics:

---

[6] See the comprehensive surveys by Christoffersen et al. (2013) and Poon and Granger (2003) or, more recently, Barone-Adesi et al. (2018) and Crisóstomo and Couso (2018). Figlewski (2018) provides a review on the use of RNDs.

[7] See, among others, Bliss and Panigirtzoglou (2004), Shackleton et al. (2010), Kostakis et al. (2011), Høg and Tsiaras (2011) and DeMiguel et al. (2013).

$$F_T = F_t e^{\lambda \tau + H(\tau; \sigma, v, \theta)}$$

$$\lambda = \frac{1}{v} \ln(1 - \theta v - \frac{\sigma^2 v}{2}) \quad \quad \quad (4)$$

$$H(\tau; \sigma, v, \theta) = \theta G(\tau; v) + \sigma G(\tau; v) W_t$$

where $G(\tau; v)$ is a Gamma distribution; $\sigma$, $v$ and $\theta$ control the volatility, skewness, and kurtosis of the distribution, and $\tau$ represents the time between the observation date $t$ and the forecast horizon $T$.

Risk-neutral distributions for all stochastic processes are obtained by inversion of their characteristic functions. Heston (1993) and Bakshi and Madan (2000) show that the cumulative distribution function (CDF) of a stochastic process $F_T$ can be obtained as:

$$CDF_T(k) = \frac{1}{2} + \frac{1}{\pi} \int_0^\infty \text{Re}\left[ \frac{e^{-iw\ln(k)} \psi_{\ln F_T}(w)}{iw} \right] dw \quad \quad \quad (5)$$

where $\psi_{\ln F_T}(w)$ denotes the characteristic function of $\ln(F_T)$, $\text{Re}[\cdot]$ is the real operator and $k$ are the strikes prices[8].

Finally, we employ the Malz (2014) implementation of the Breeden and Litzenberger (1978) formula (BLMALZ). Malz's approach generates an implied volatility function using cubic spline interpolations across the available implied volatilities and a flat extrapolation at endpoints. For each forecast date, the implied volatility function is employed to compute a continuous of call prices $C(k, \tau)$, which is numerically differentiated to obtain the CDF which replicates option prices:

$$CDF_T(k) \approx 1 + e^{r_t \tau} \frac{1}{\Delta}\left[ C(k - \frac{\Delta}{2}, \tau) - C(k + \frac{\Delta}{2}, \tau) \right] \quad \quad \quad (6)$$

where $\Delta$ denotes the step size for the finite differentiation.

## 2.2 Risk-adjusted densities

We next examine how RNDs can be adjusted to incorporate investor's risk preferences. Following Cochrane (2005), the present value of any asset $P_t$ can be computed as the expectation of future cash flows, discounted by the SDF.

$$P_t = E_t\left[ m_T z_T \right] \quad \quad \quad (7)$$

where $E_t$ is the expectation under the representative investor measure, $z_T$ are the state-dependent payoffs, and $m_T$ is the pricing kernel which summarizes investor's preferences and beliefs. Expressed in integral form, the pricing equation becomes:

$$P_t = \int_\mathbb{R} m_t(x_T) z_T(x_T) f_t(x_T) dx_T \quad \quad \quad (8)$$

---

[8] See Crisóstomo (2014, 2017) for further details.

where $f_t(x_T)$ represents the probability density function of future asset prices $x_T$. However, since $f_t$ does not yet include the impact of investor's preferences, which are embedded in $m_T$, these probabilities cannot be deemed as representative of investor's expectations for future prices.

In a risk-neutral world, investor's preferences reflect time-discount only. Therefore, the pricing kernel is simply $m_T^{\mathbb{Q}} = e^{-r\tau}$, and by substituting in (8), the value of any asset becomes:

$$P_t = e^{-r\tau} \int_{\mathbb{R}} z_T(x_T) f_t^{\mathbb{Q}}(x_T) dx_T \qquad (9)$$

where $f_t^{\mathbb{Q}}(x_T)$ represents the risk-neutral density function calculated at $t$ and with forecast horizon $T$. Since $m_T^{\mathbb{Q}}$ contains the discount-factor only, which does not vary across wealth states, the risk-neutral SDF does not modify the generic probabilities in (8), and thus $f_t^{\mathbb{Q}} \equiv f_t$.

In contrast, in a risk-adjusted economy, rational investors exhibit risk-aversion, attaching a decreasing marginal utility to payoffs received in states of higher wealth. Liu et al. (2007) show that the pricing kernel for risk-averse investors is proportional to its marginal utility and given by $m_T^{RA} = e^{-r\tau} u'(x_T)$. Therefore, the pricing equation becomes:

$$P_t = e^{-r\tau} \int_{\mathbb{R}} z_T(x_T) f_t^{RA}(x_T) u'(x_T) dx_T \qquad (10)$$

Comparing (9) and (10), it follows that for the risk-neutral and risk-adjusted economies to generate the same market price, the risk-adjusted density $f_t^{RA}$ should be equal to $f_t^{\mathbb{Q}}/u'$. However, Bliss and Panigirtzoglou (2004) note that the transformation from risk-neutral to risk-adjusted probabilities exhibits non-linearities, and hence the scaling factor $\int_o^\infty f_t^{\mathbb{Q}}(y)/u'(y) dy$ should be employed to ensure integration to unity. Consequently, starting from any RND, the risk-adjusted probabilities can be obtained as:

$$f_t^{RA}(x_T) = \frac{f_t^{\mathbb{Q}}(x_T)/u'(x_T)}{\int_o^\infty f_t^{\mathbb{Q}}(y)/u'(y) dy} \qquad (11)$$

In particular, for investors featuring a power utility with constant relative risk aversion (CRRA),

$$u^{CRRA}(x_T) = \begin{cases} \frac{1}{1-\gamma} x_T^{1-\gamma} & \text{if } \gamma \geq 0; \gamma \neq 1 \\ \ln x_T & \text{if } \gamma = 1 \end{cases} \qquad (12)$$

where $\gamma$ denotes the coefficient of relative risk-aversion. The marginal CRRA utility is then $u' = x_T^{-\gamma}$, and the pricing kernel becomes $m_T^{CRRA} = e^{-r\tau} x_T^{-\gamma}$. Therefore, the risk-adjusted density $f_t^{CRRA}$ is:

$$f_t^{CRRA}(x_T) = \frac{x_T^\gamma f_t^{\mathbb{Q}}(x_T)}{\int_o^\infty y^\gamma f_t^{\mathbb{Q}}(y) dy} \qquad (13)$$

In addition, when $\gamma = 0$, the CRRA pricing kernel is $m_T^{CRRA} = e^{-r\tau}$, and thus the CRRA utility contains as special case an economy where investors are risk-neutral.

## 2.3 Real-world densities

We next consider how to embody investor sentiment in the SDF framework. To comprehensively examine the impact of behavioral effects in different areas of the return distribution, we aggregate three sentiment-induced mistakes: excessive optimism, which relates to biases in average returns; overconfidence, which leads to errors in volatility predictions, and tail sentiment, which is linked to non-rational tail expectations.

### 2.3.1 Investor optimism and overconfidence

For any density forecast $f_t(x_T)$, the behavioral transformation due to investor optimism and overconfidence can be obtained through a linear mapping of the original values $x_T$ into the adjusted values $\hat{x}_T$ [9]

$$\hat{x}_T = \theta_{1,t} + \hat{x}_T \theta_{2,t} + (1-\theta_{2,t})\mu_{x_T} \qquad (14)$$

where $\theta_1$ and $\theta_2$ denote the location and scale shift parameters, respectively. This transformation shifts the mean and standard deviation of the traditional forecast into the adjusted values $\hat{\mu} = \mu + \theta_1$ and $\hat{\sigma} = \sigma \theta_2$, providing a flexible way to adjust density predictions.

When market prices are driven by pessimist investors, market-based densities reflect a downward bias in average returns; thus, we perform a behavioral adjustment $\theta_1 > 0$ that shifts probability mass towards more favorable outcomes. Conversely, when prices are dominated by optimistic investors, traditional forecasts are corrected through a behavioral shift $\theta_1 < 0$ that moves the market-implied distribution to the left, reducing the mean.

Similarly, when prices are driven by underconfident investors, excessive dispersion is embedded in market-implied forecasts and hence a behavioral correction $\theta_2 < 1$ reduces the volatility of the distribution. In contrast, if prices are driven by overconfident investors, an adjustment $\theta_2 > 1$ increases the volatility of the market-implied forecast.

After the behavioral transformation, the mean-variance (mv) pricing kernel $m_T^{mv}$ is obtained as:

$$m_T^{mv} = \frac{f_t^{RA}(x_T)}{f_t^{mv}(x_T)} \qquad (15)$$

where $f_t^{RA}(x_T)$ represents the risk-adjusted density and $f_t^{mv}(x_T)$ is the behavioral distribution obtained through the mean-variance shift.

---

[9] A downward sloping SDF shifts probability mass from left to right, generating a transformation similar to an upward mean change. Likewise, a U-shaped pricing kernel shifts probability from tails to center, producing a downward-like volatility move. However, for the variety of models used in this study, there is no simple SDF which can generate a customary mean-variance shift.

### 2.3.2 Tail sentiment

To account for biases in tail expectations, we employ a simple adjustment which progressively shifts probability mass from the left to the right tail, or vice versa. Denoting by $q(\alpha)$ and $q(1-\alpha)$ the quantiles which define the left and right tails respectively, we obtain the tail-shift pricing kernel $m_T^{ts}$ as

$$m_T^{ts} = \begin{cases} e^{\theta_3(q(\alpha)-x_T)} & \text{for } x_T \in (-\infty, q(\alpha)) \\ 1 & \text{for } x_T \in [q(\alpha), q(1-\alpha)] \\ e^{-\theta_3(x_T-q(1-\alpha))} & \text{for } x_T \in (q(1-\alpha), \infty) \end{cases} \quad (16)$$

where $\theta_3$ controls the direction and intensity of the tail-shifting. When excessive left-tail fear is embedded in market prices, investors overstate left tail events compared to the right tail; hence a behavioral transformation $\theta_3 > 0$ progressively shifts probability from the left to the right tail. Conversely, if investor exhibits right-tail exuberance, an adjustment $\theta_3 < 0$ increases the severity of the left tail, reducing the right tail. The log-linearity of $m_T^{ts}$ ensures that all tail-adjusted probabilities remain positive even for extreme values of the density domain.

### 2.3.3 Sentiment function and real-world SDF

The sentiment function $\Psi(x_T)$ summarizes the behavioral corrections required to transform risk-adjusted forecasts into real-world densities. In terms of the SDF, the aggregate effect of investor optimism, investor overconfidence and tail sentiment is obtained as:

$$\Psi(x_T) = m_T^{mv} m_T^{ts} \quad (17)$$

Consequently, the real-world pricing kernel $m_T^{RW}$, which reflects the cumulative impact of investor sentiment and investor risk preferences is given by:

$$m_T^{RW} = e^{-r\tau} u'(x_T) \Psi(x_T) \quad (18)$$

Intuitively, a real-world SDF of 1.2 for an *ex-post* return $\tilde{x}_T$, indicates that the traditional forecast overestimates the likelihood of $\tilde{x}_T$ by a factor of $1.2/e^{-r\tau}$. More formally, following equations (9) to (11), the real-world density can be obtained from an initial RND $f_t^{\mathbb{Q}}$, marginal utility $u'$, and sentiment function $\Psi$ as:

$$f_t^{RW}(x_T) = \frac{f_t^{\mathbb{Q}}(x_T)/u'(x_T)\Psi(x_T)}{\int_o^\infty f_t^{\mathbb{Q}}(y)/u'(y)\Psi(y)dy} \quad (19)$$

where the scaling factor $\int_o^\infty f_t^{\mathbb{Q}}(y)/u'(y)\Psi(y)dy$ is used to ensure integration to unity.

### 2.4 Practical Illustration

The sentiment function $\Psi(x_T)$ provides a flexible way to transform traditional predictions into real-world forecasts. Figure 1 shows three real-world densities that have been obtained under different calibrations of the sentiment tuple $\{\theta_1, \theta_2, \theta_3\}$. In all cases, the initial density is lognormal and investor risk preferences are modeled through a CRRA utility with $\gamma = 2$.

**Figure 1: From traditional to real-world distributions**

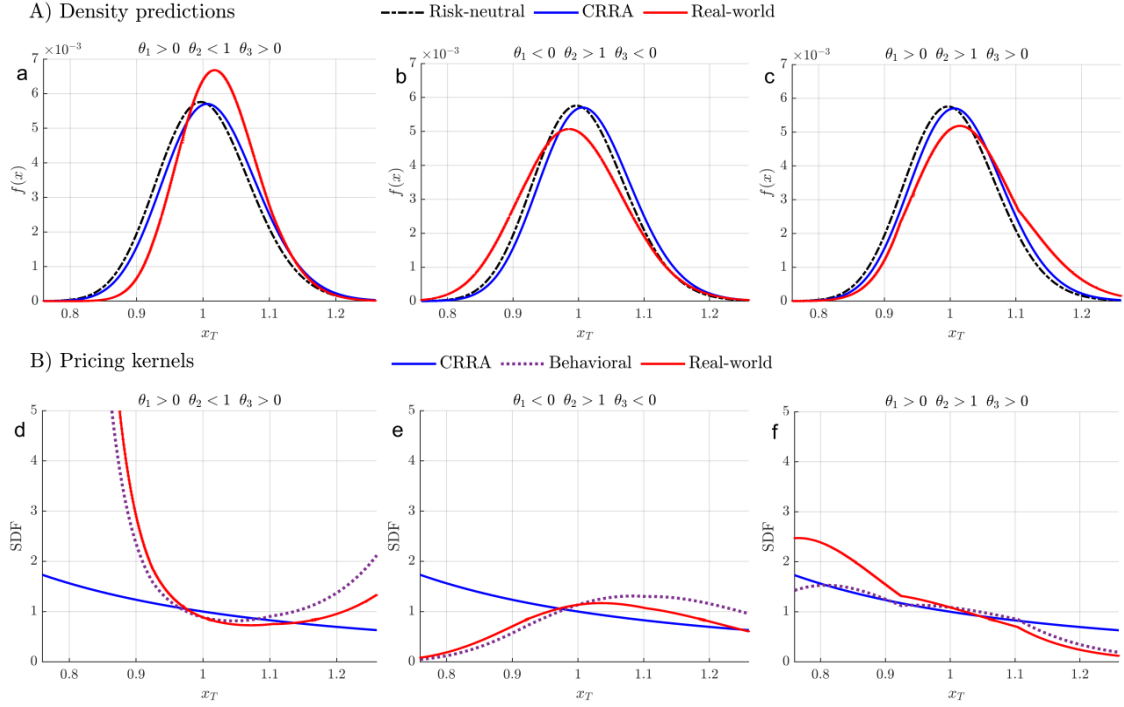

Notes: Figures 1a to 1c exhibit the transformation from risk-neutral into risk-adjusted and real-world distributions. Figures 1d to 1c show the pricing kernels used in each transformation. The CRRA and real-world densities are obtained by dividing the RND by the CRRA and real-world pricing kernels, respectively (excluding the discount factor $e^{-r\tau}$). Alternatively, real-world densities can be also obtained by dividing the CRRA density by the behavioral SDF. The CRRA kernel is $m_T^{CRRA} = e^{-r\tau} x_T^{-\gamma}$ and the real-world kernel is $m_T^{RW} = e^{-r\tau} x_T^{-\gamma} \Psi(x_T)$, where $\Psi(x_T)$ represents the behavioral SDF or sentiment function. The sentiment parameters used in the transformation $\{\theta_1, \theta_2 \text{ and } \theta_3\}$ are shown above each figure.

Figure 1a depicts a characterization that is representative of investor's fear. Market prices are driven by pessimism and underconfident investors which overstate the left tail. Consequently, our sentiment correction shifts probability mass towards more positive outcomes ($\theta_1 > 0$), reduces the volatility of the distribution ($\theta_2 < 1$), and increases the right tail ($\theta_3 > 0$), generating a real-world density that is concentrated around more favorable outcomes.

Conversely, Figure 1b presents an initial forecast that is characterized by investor's complacency. Asset prices are driven by optimist and overconfident investors which assign low probabilities to the left tail. Therefore, the behavioral density is displaced to the left ($\theta_1 < 0$), exhibits higher overall dispersion ($\theta_2 > 1$) and shows an increased left tail ($\theta_3 < 0$).

Finally, the market-implied distribution in Figure 1c is driven by a mix of optimist and overconfident investors and those that assign too much probability to the left tail. After the sentiment correction, the real-world density exhibits increased dispersion ($\theta_2 > 1$) and higher probabilities at the right end of the distribution ($\theta_1 > 0$ and $\theta_3 > 0$). Conversely, the effects of overconfidence and left-tail fear partially offset each other towards the left side, reducing the impact of the behavioral shift.

The pricing kernels used in each transformation are depicted in Figures 1d to 1f. First, all risk-neutral forecasts are adjusted to incorporate investor`s risk preferences through the CRRA kernel. Intuitively, for SDF values above (below) 1, the initial density overestimates (underestimates) the risk-adjusted probabilities. Therefore, the monotonically decreasing CRRA kernel shifts probability towards more favorable outcomes, generating a transformation resembling an upward mean change.

Risk-adjusted densities are transformed into real-world forecasts using the behavioral SDF calculated through the calibrated parameters $\{\theta_1, \theta_2 \text{ and } \theta_3\}$. In figure 1d, overconfidence is the main driver of the U-shaped sentiment function. However, the slope of the pricing kernel is higher at the left side. The asymmetry can be explained by the corrections for investor pessimism and left-tail fear, which shift probability away from negative outcomes, triggering a higher SDF increase. In contrast, the correction for overconfidence partially offsets the probability gains stemming from investor pessimism and left-tail fear over the right side, limiting the pricing kernel increase.

In Figure 1e, investor underconfidence produces a generally inverse U-shaped behavioral SDF. In addition, the corrections for optimism and right-tail exuberance shift probability mass from positive to negative outcomes, leading to a higher SDF decrease at the left tail. The combined effect generates a sentiment function that shrinks rapidly towards the left end of the distribution, stays above 1 for moderately positive outcomes, and slowly decreases again towards the right tail.

Finally, in Figure 1f the mix of investor pessimism, overconfidence and left-tail fear generate an oscillating sentiment function that resembles the behavioral SDF in Shefrin (2008). The behavioral SDF is generally decreasing but exhibits an upward-sloping part near the center of the distribution, reproducing the pricing kernel puzzle. The decreasing SDF is driven by investor pessimism and left-tail fear, both shifting probabilities towards more favorable prices. In contrast, the correction for overconfidence reduces the probability at the center of the distribution, increasing the pricing kernel in this region and generating an upward-sloping section in the sentiment function.

## 3. Measuring sentiment effects

A challenge for researchers is that investor sentiment is not directly observable. Sentiment effects manifest through asset prices, but there is no consensus on which methodology and variables should be used to disentangle sentiment from fundamental expectations. Among the possible choices, our approach to measuring investor sentiment is threefold:

First, we obtain our sentiment proxies directly from market activity. The underlying reason is that it is the day-to-day activity of market participants which impacts asset prices and generate biases in market expectations. In contrast to investor surveys or social media, market-based inputs stem from the decisions taken by active investors that are currently trading in the markets. Furthermore, the rise of algorithmic trading underscores the use of market-based inputs, as algorithmic trading represents a growing market portion but seldom appears in surveys or social media.

Second, we estimate behavioral effects using forward-looking inputs that are updated daily. This contrasts with the use of proxies that are computed on a low frequency or published with a delay over the measurement period, hence exhibiting time lags and a limited ability to capture day-to-day changes in investor's beliefs.

Third, empirical research suggests that behavioral biases manifest in asset prices specifically during high sentiment periods.[10] Heightened sentiment phases attract more noise traders, increasing the proportion of sentiment-induced trades and undermining market efficiency. Therefore, instead of correcting density predictions at all possible dates, our framework specifically links pricing anomalies to high sentiment periods.

### 3.1 Investor optimism

Our proxy for investor optimism is derived from implied volatility extracted from option prices. When investors are pessimistic about future returns, they increase demand for hedging securities, rising implied volatilities. Empirically, several papers show implied volatility to be a contrarian indicator of average returns, concluding that high (low) implied volatilities predict lower (higher) subsequent returns[11]. Furthermore, Smales (2017) shows that changes in implied volatility, rather than overall levels, are preferable to explain future returns.

For each observation date, we calculate the implied volatility change $\Delta IV_t$ as the difference between the current implied volatility $IV_t$ and the average implied volatility in the previous three months $\overline{IV}_{t-1} = \sum_{s=1}^{3} IV_{t-s} / 3$, hence:

$$\Delta IV_t = IV_t - \overline{IV}_{t-1} \qquad (20)$$

The set of available volatility changes up to date $t$ $\{\Delta IV_1, ..., \Delta IV_{t-1}\}$ is transformed into a continuous distribution through a Gaussian kernel with bandwidth estimated with the formula of Silverman (1996). Each new $\Delta IV_t$ observation is mapped into a quantile $\alpha_t^{IV}$ of the corresponding empirical distribution. Observations falling below the 5$^{th}$ quantile are associated with excessive optimism, whereas those beyond the 95$^{th}$ quantile are linked to excessive pessimism. The behavioral mean-shift $\theta_{1,t}$ is calibrated as:

$$\theta_{1,t} = \begin{cases} (1-e^{r_t \tau})k_1 \frac{0.05-\alpha_t^{IV}}{0.05} & \text{for} \quad 0 < \alpha_t^{IV} < 0.05 \\ 0 & \text{for } 0.05 \leq \alpha_t^{IV} \leq 0.95 \\ -(1-e^{r_t \tau})k_1 \frac{\alpha_t^{IV}-0.95}{0.05} & \text{for } 0.95 < \alpha_t^{IV} < 1 \end{cases} \quad (21)$$

---

[10] Similar to Bondt et al. (2015), we do not allege that market efficiency is broken, only that it can break. Studies that link high sentiment periods and pricing anomalies include, among others: Yu and Yuan (2011), Stambaugh et al. (2012), Shen et al. (2017), and Lin et al. (2018).

[11] See for instance: Whaley (2009), Simon and Wiggins (2001), Giot (2009), and Smales (2017).

To gauge how different calibrations impact probabilistic forecasts, we test two alternative values for $k_1$. In our low impact calibration, investor optimism may lead to a mean-shift of up to one time the risk-free rate ($k_1 = 1$), whereas in the high impact case the behavioral shift is twice as much ($k_1 = 2$).

## 3.2 Overconfidence

Overconfidence is the tendency to place an excessive degree of confidence in one's abilities and beliefs (Grinblatt and Keloharju, 2009). Investor overconfidence can be proxied by the volume traded in the market. The link between overconfidence and trading volumes is well established in the literature, showing that overconfident investors overstate the precision of their private information, understating future volatility and trading more than rational models suggest.[12]

Similar to investor optimism, Statman et al. (2006) find that changes in trading volumes rather than overall levels are preferable to measure overconfidence. On each date $t$, we compute the change in traded volume $\Delta TV_t$ as the ratio of the last month volume $TV_t$ compared to the average volume in the previous three months $\overline{TV}_{t-1} = \sum_{s=1}^{3} TV_{t-s} / 3$, thus:[13]

$$\Delta TV_t = TV_t / \overline{TV}_{t-1} \quad (22)$$

Next, the set of volumes changes $\{\Delta TV_1, ..., \Delta TV_{t-1}\}$ is transformed into a continuous distribution using a Gaussian KDE. The volatility adjustment $\theta_{2,t}$ is then obtained as:

$$\theta_{2,t} = \begin{cases} k_2^{-1} \frac{0.05-\alpha_t}{0.05} & \text{for} \quad 0 < \alpha_t^{TV} < 0.05 \\ 0 & \text{for } 0.05 \leq \alpha_t^{TV} \leq 0.95 \\ k_2 \frac{\alpha_t-0.95}{0.05} & \text{for } 0.95 < \alpha_t^{TV} < 1 \end{cases} \quad (23)$$

where $\alpha_t^{TV}$ represent the quantile position in the empirical distribution. $\Delta TV_t$ observations falling below the 5$^{th}$ quantile of the distribution are considered as underconfidence, whereas those above the 95$^{th}$ are associated with overconfidence. In the low impact scenario $k_2$ = 1.2 and hence the volatility multiplier $\theta_{2,t}$ ranges from 0.83 to 1.2, whereas in the high sentiment case $k_2$ = 1.5 and the volatility adjustment varies from 0.67 to 1.5.

---

[12] See among others: De Bondt and Thaler (1995), Odean (1999), Barber and Odean (2001), Glaser and Weber (2007), Grinblatt and Keloharju (2009), Statman, Thorley, and Vorkink (2006), Abreu and Mendes (2012) and Michailova and Schmidt (2016)

[13] Given the structural increase in trading volumes during the sample period, we derive our proxy for overconfidence from short-term volumes changes, hence avoiding the impact of long-term trends.

### 3.3 Tail sentiment

Tail sentiment is related to the way in which investors price and perceive tail risk. Since market participants typically live in fear of a crash in prices, they buy OTM puts beyond what rational levels suggest, contributing to a sentiment-induced overpricing in OTM puts[14].

The amount of tail risk priced in the market can be inferred from the skewness of the risk-neutral distribution. When investors perceive substantial downside risks, the risk-neutral distribution becomes more negatively skewed. In contrast, when investors perceive that an asset offers high returns and limited risks, its risk-neutral skewness becomes positive. On each date $t$, we obtain the RND skewness through the model-free method of Bakshi et al. (2003):

$$Skew_t = \frac{e^{r\tau}W(t,T) - 3\mu(t,T)e^{r\tau}V(t,T) + 2\mu(t,T)^3}{(e^{r\tau}V(t,T) - \mu(t,T)^2)^{3/2}} \quad (24)$$

where

$$\mu(t,T) = e^{r\tau}\left(1 - e^{-r\tau} - \frac{1}{2}V(t,T) - \frac{1}{6}W(t,T) - \frac{1}{24}X(t,T)\right) \quad (25)$$

and $V(t,T)$, $W(t,T)$ and $X(t,T)$ represent hypothetical contracts with a quadratic, cubic and quartic payoff respectively, and are obtained as described in Bakshi et al. (2003).

Following empirical evidence, we examine whether high skewness levels can be associated with biases in tail expectations[15]. Specifically, we consider that sentiment biases are embodied in asset prices when $Skew_t$ exceeds a ± 1.5 threshold. Since equity distributions typically exhibit negative skewness, this choice is expected to generate more frequent overweightings of the left tail compared to the right rail, which is consistent with existing evidence. The tail-shift adjustment $\theta_{3,t}$ is then obtained as:

$$\theta_{3,t} = \begin{cases} k_3(Skew_t + 1.5) & for \ Skew_t < -1.5 \\ 0 & for \ -1.5 \leq Skew_t \leq 1.5 \\ k_3(Skew_t - 1.5) & for \ Skew_t > 1.5 \end{cases} \quad (26)$$

In our low sentiment calibration the increase in $\theta_{3,t}$ is linear ($k_3$ = 1), whereas in the high sentiment case the increase is twice as much ($k_3$ = 2).

---

[14] See Bondarenko (2014), Schreindorfer (2014), Greenwood-Nimmo, Nguyen, and Rafferty (2016) and Bollerslev and Todorov (2011). Note that investor's perception of tail risk shifts over time and market participants may also underestimate left-tail risks.

[15] See Han (2008), Chen and Gan (2018) and Bevilacqua et al. (2018).

## 4. Data and Calibration

Our data selection process maximizes the use of market prices and minimizes the input modeling assumptions, thus following the recommendations in Christoffersen et al. (2013). Moreover, all model parameters are estimated strictly out-of-sample, using only information know to investors up to time $t$.

### 4.1 Option data

Our database is comprised of European-style options with underlying the IBEX 35 futures. Option prices are retrieved from the official Spanish derivatives exchange (MEFF) from November 1995 to December 2016, covering over 21 years. Our study focuses on market-derived option prices only. This choice contrasts with the use of exchange-reported settlement prices, which in many cases are theoretically estimated and do not reflect real trading activity[16].

To maximize input's representativeness and sample size, we work with front-month option contracts; options with monthly maturity are the most actively traded in MEFF and maximize the number of non-overlapping periods[17]. Observation dates are set 28 calendar days before each expiry date. On each date $t$, we record the price of all available call and put options exhibiting contemporaneous bid and ask quotes. Since ITM options are less actively traded than OTM options, we build our dataset with OTM and ATM options. To evaluate the consistency of each cross-section, OTM put are converted into call equivalent prices using the Put-Call parity. Options that do not respect non-arbitrage conditions are removed from the dataset[18].

After filtering, we obtain 6659 option prices distributed across 254 monthly cycles. The average number of strikes in the cross-sections is 26, ranging from a minimum of 8 to a maximum of 72. Table 1 summarizes the statistics of the option dataset.

Separately, the final settlement price for each monthly future is determined by averaging the IBEX 35 spot prices from 16:15 to 16:45. These settlement prices constitute the underlying asset of the IBEX 35 options and futures used in this study, and are hence used to assess the forecasting ability of each predictive model.

---

[16] For instance, the daily settlement prices of IBEX 35 options are theoretically computed by MEFF assuming a linear relation in the implied volatility function for OTM and ITM options. Therefore, these settlement prices reflect specific modeling choices and using them in the calibration would entail introducing an exogenously derived volatility shape.

[17] IBEX 35 options do not exhibit shorter than monthly expirations. Therefore, shorter consecutive periods would lead to forecast horizons that lack direct market quotes, requiring extrapolation assumptions. On the other hand, longer expiration cycles would entail reducing the number of non-overlapping periods and relying on less liquid back-month contracts.

[18] Call and put-derived equivalent contracts whose price is not a convex and decreasing function of the strike are removed from the dataset.

**Table 1: Summary statistics for the option dataset**

| Option type | Total number | Average per day | Maximum per day | Minimum per day |
|---|---|---|---|---|
| Calls | 3151 | 12 | 38 | 1 |
| Puts | 3508 | 14 | 46 | 3 |
| Overall | 6659 | 26 | 72 | 8 |

| Moneyness | F/K | No. of options | (%) |
|---|---|---|---|
| Deep OTM put | >1.10 | 1755 | 26.36 |
| OTM put | 1.03-1.10 | 1423 | 21.37 |
| Near the money | 0.97-1-03 | 1496 | 22.47 |
| OTM call | 0.90-0.97 | 1541 | 23.14 |
| Deep OTM call | <0.90 | 444 | 6.67 |

Notes: A minimum of 8 options is required to calibrate the Bates model. Therefore, in seven observation dates, we supplemented the cross-sections with at-the-money option contracts whose last traded price was consistent with the contemporaneous bid and ask prices. This resulted in an addition of 9 options (0.1% of the sample)

### 4.2 Interest rates and dividends

From January 1999 to December 2016, we employ the 1-month Euribor. On earlier dates, since the Euribor was not yet available, we employ the 1-month Mibor. Effective interest rates are computed for each forecasting period using the corresponding act/360 day count convention. The use of futures contracts makes dividend estimation irrelevant; hence, dividend uncertainties do not affect our density forecasts.

### 4.3 Sentiment data

Our sentiment framework avoids information mismatches through the use of consistent forward-looking inputs. On each date $t$, investor optimism and tail sentiment are derived from the cross-section of available option prices, matching the expectations embedded in option-implied densities. Similarly, overconfidence is extracted from changes in traded volumes that are updated at each forecast date, hence reflecting forward-looking expectations and up-to-date investor's beliefs.

Our proxy for investor optimism is derived from ATM volatilities. On each monthly date from December 1994-2016, the ATM volatility $IV_t^{atm}$ is computed by linear interpolation of the two closest-to-the-money options in each cross-section. Next, we obtain the change compared with the three previous months, $\Delta IV_t^{atm}$, which is employed to estimate our optimism correction $\theta_{1,t}$.

To estimate overconfidence, trading volumes in the IBEX 35 are collected from November 1994 to December 2016. Trading volumes at each forecast date are computed as the cumulative volume over the most recent 20 business days. The ratio of the current volume compared with the average volume in the three previous months $\Delta TV_t$ is used as proxy for overconfidence and employed to compute the overconfidence correction $\theta_{2,t}$ [19].

Finally, tail sentiment is derived from the RND skewness embedded in option prices. For each observation date, we obtain $Skew_t$ through the model-free method of Bakshi et al. (2003) and skewness levels are used to calculate the behavioral tail corrections $\theta_{3,t}$.

**4.4 Calibration of traditional densities**

Consistent with sentiment effects, all traditional densities are estimated strictly out-of-sample, using the information available up to date $t$. Risk-neutral forecasts are derived from the cross-section of option prices. For the BSM model, the ATM volatilities $IV_t^{atm}$ and risk-free rates $r_t$ are used to obtain the corresponding lognormal densities. For the Heston, Bates and Variance Gamma models, we estimate the parameter set $\hat{\Theta}_t$ that minimizes the sum of relative errors:

$$SRE_t = \sum_{i=1}^{N_t} \frac{|C_i - \hat{C}_i(\tilde{\Theta}_t)|}{C_i} \quad (27)$$

where $N_t$ is the number of option prices available at date $t$, $C_i$ the mid-market price of each option in the cross-section and $\hat{C}_i(\tilde{\Theta})$ is the model-dependent value obtained with the parameter set $\tilde{\Theta}$ [20]. Following this procedure, risk-neutral parameters are individually estimated for each stochastic model, hence performing 254 calibrations per model. For Malz's implementation of Breeden-Litzenberger formula, we fix the step size for the finite differences at $\Delta = 0.01 F_t$, which avoids negative probabilities in our density forecasts.

Regarding risk preferences, Cuesdeanu and Jackwerth (2018a) note that investor preferences are notoriously difficult to estimate; risk aversion estimates tend to be unstable and change widely over short periods. Barone-Adesi et al. (2017) argue that this problem could be linked to the traditional way of estimating risk preferences which, by ignoring sentiment, forces investor biases to manifest through the risk aversion parameter.

To obtain consistent risk-aversion estimates, we follow the forward-looking approach in Cuesdeanu and Jackwerth (2018b), which employs a power utility function with CRRA coefficients of 0, 2 and 4. This choice is supported by both empirical and theoretical evidence. Thomas (2016) finds that the power utility family is the only valid utility class and commonly used in practical applications. Furthermore, our CRRA coefficients are among the most employed in the forecasting literature[21]. Section 6.6 further expands these analyses with an alternative risk setting where risk aversion is dynamically estimated from option prices.

---

[19] Since trading volumes in August exhibit a cyclical decrease that is not related to investor confidence, we adjust August figures through the ratio of the volume traded in past August months compared to the volume in the three previous months.
[20] The use of relative errors assigns a similar weight to all option contracts, thus generating consistent results across different strike regions. See Heston (1993), Bates (1996) and Madan et al. (1998) for a description of the admissible parameter values. In the VG process, we also consider the restriction $v^{-1} > \theta + \sigma^2/2$, which avoids numerical calibration problems (see Itkin, 2010 and Crisóstomo, 2017).
[21] See Bliss and Panigirtzoglou (2004), Meyer and Meyer (2005), Polkovnichenko and Zhao (2013), Barone-Adesi et al. (2017) and Brinkmann and Korn (2018).

## 5. Forecast evaluation

We evaluate the predictive ability of all density forecasts using three complementary criteria. First, we analyze the log-likelihood of each predictive density with the logarithmic score. Second, forecast errors are assessed through the CRPS. Third, statistical consistency is examined by means of the Berkowitz, Jarque-Bera and Kolmogorov-Smirnov goodness-of-fit tests. Finally, all density models are ranked in a standardized scale using the IFS.

### 5.1 Logarithmic score

The accuracy of different probabilistic forecasts can be compared through the likelihood of the *ex-post* realizations evaluated with the *ex-ante* densities. Following Liu et al. (2007), Shackleton et al. (2010) and Høg and Tsiaras (2011), we obtain the logarithmic score for each forecast model as:

$$L = \sum_{t=1}^{N} \log(f_t(\tilde{x}_T)) \tag{28}$$

where $f_t$ denotes the density forecast computed at time $t$, $\tilde{x}_T$ denotes the *ex-post* realization at $T$, and $N$ is the number of non-overlapping forecasts. By aggregating the scores over the entire sample, forecast models can be ranked in terms of their out-of-sample accuracy. Furthermore, Bao, Lee, and Saltoğlu (2007) show that, when all models are possibly misspecified, as is the case in financial forecasts, the predictive scheme with the highest $L$ is nearest to the true generating density in terms of the Kullback-Leibler Information Criterion.

### 5.2 Continuous Ranked Probability Score

The CRPS evaluates the entire predictive distribution, measuring the distance between the *ex-post* realization and all the *ex-ante* probability masses (Matheson and Winkler, 1976). As a result, the CRPS gives good scores to densities that assign high probabilities to values that are close but are not identical to the one materializing (Gneiting and Raftery, 2007), complementing the local accuracy angle provided by the log score.

Denoting by $CDF_t^m$ and $CDF_t^r$ the cumulative distributions of the forecasting model and the realization, the CRPS for a probabilistic prediction is:

$$CRPS_t = \int_{-\infty}^{\infty} \left( CDF_t^m(x) - CDF_t^r(x) \right)^2 dx \tag{29}$$

where:

$$CDF_t^r(x) = \begin{cases} 0 & for \ \tilde{x}_T < x \\ 1 & for \ \tilde{x}_T \geq x \end{cases} \tag{30}$$

The CRPS has the dimension of parameter $x$, which enters in the calculus through $dx$, facilitating the interpretation of the CRPS as a generalization of the mean absolute error (Hersbach, 2000). To ensure the comparability across different periods, we calculate the CRPS using return deviations instead of index points, computing the average CRPS for the entire sample as:

$$CRPS = \frac{1}{N}\sum_{t=1}^{N}\sqrt{\int_{-\infty}^{\infty}\left(CDF_t^m(x) - CDF_t^r(x)\right)^2 dx} \tag{31}$$

For a given model, a $CRPS$ of 0.03 can be interpreted as an average return deviation of 3% between the *ex-post* realizations and the *ex-ante* probabilistic outcomes, providing an intuitive metric to rank density predictions.

**5.3 Goodness-of-fit tests**

Following Diebold et al. (1998) the statistical consistency of an ensemble of forecasts can be assessed through PIT-based analyses. For a given date $t$, the PIT represents the quantile of the *ex-ante* distribution at which the *ex-post* realization is observed. Thus,

$$PIT_t = \int_{-\infty}^{\tilde{x}_T} f_t(x)\, dx \tag{32}$$

In a correctly specified model, quantile realizations should be indistinguishable from random draws from the predictive distributions. Consequently, the sequence of PIT values should converge to a uniform distribution. However, tests based on uniform variables are typically not powerful enough for small samples. Berkowitz (2001) modifies the PIT realizations into a transformed sequence $T\text{-}PIT_t = \Phi^{-1}(PIT_t)$ that should be formed by $N(0,1)$ i.i.d. variables if the density is correctly specified. The Berkowitz test jointly assesses the mean, variance and independence of the T-PIT sequence:

$$T\text{-}PIT_t - \mu = \rho(T\text{-}PIT_{t-1} - \mu) + \varepsilon_t \tag{33}$$

through the likelihood ratio test LR3 = $-2(L(0,1,0) - L(\hat{\mu}, \hat{\sigma}^2, \hat{\rho}))$, which compares the likelihood of a restricted model [where $\mu = 0$, var($\varepsilon_t$) = 1 and $\rho = 0$] with that of an unrestricted one.

Nevertheless, as indicated in Dowd (2004), the Berkowitz test does not specifically assess the normality of the T-PIT sequence, hence being unable to detect failures in higher moments. To complement the LR3, we employ the Jarque-Bera (JB) and Kolmogorov-Smirnov (KS) tests. The KS test examines whether the maximum distance between the T-PIT distribution and a $N(0,1)$ is statistically significant, whereas the JB test specifically assesses the skewness and kurtosis of the T-PIT realizations.

**5.4 Integrated Forecast Score**

The results from the log score, CRPS and goodness-of-fit analyses provide complementary angles to evaluate probabilistic forecasts. However, as different metrics may lead to diverging model choices, we summarize all forecast measures in a standardized ranking using the IFS (Crisóstomo and Couso, 2018).

To compute the IFS, we first obtain the normalized [0, 1] scores for local accuracy $\overline{L}$, global errors $\overline{CRPS}$ and statistical consistency $\overline{Stat}$. The first two are derived from the normality of the overall log-likelihood and CRPS figures. Specifically, since entire sample figures are obtained from a sum of 254 independent and similarly distributed observations, they should converge to a Normal distribution. Consequently, we obtain the mean and standard deviation of the overall log-likelihood and CRPS figures, and each model is ranked in a [0, 1] scale according to its quantile position in the corresponding distribution.

To calculate the normalized score for statistical consistency, forecast models are first ranked by the number of tests passed, with 0.25 points allocated for every non-rejected test at 5% significance level. The remainder 0.25 further discriminates across competing models through the specific *p*-values achieved in the Berkowitz, JB and KS tests. For each statistical test, the highest and lowest *p*-values are assigned 0 and 1 values, whereas other *p*-values are ranked through linear interpolation.

Finally, the IFS for each predictive method is obtained by averaging the three normalized scores:

$$IFS = (\overline{L} + \overline{CRPS} + \overline{Stat})/3 \tag{34}$$

## 6. Empirical results

This section discusses the sentiment adjustments obtained through our behavioral framework and compares the out-of-sample performance of all risk-neutral, risk-adjusted and real-world densities.

### 6.1 Sentiment adjustments

Figure 2 reports the dates on which our sentiment adjustments were activated and their associated values. To better understand our corrections, Figure 2 also shows the evolution of the IBEX 35 during the sample period.[22]

**Figure 2: Sentiment adjustments**

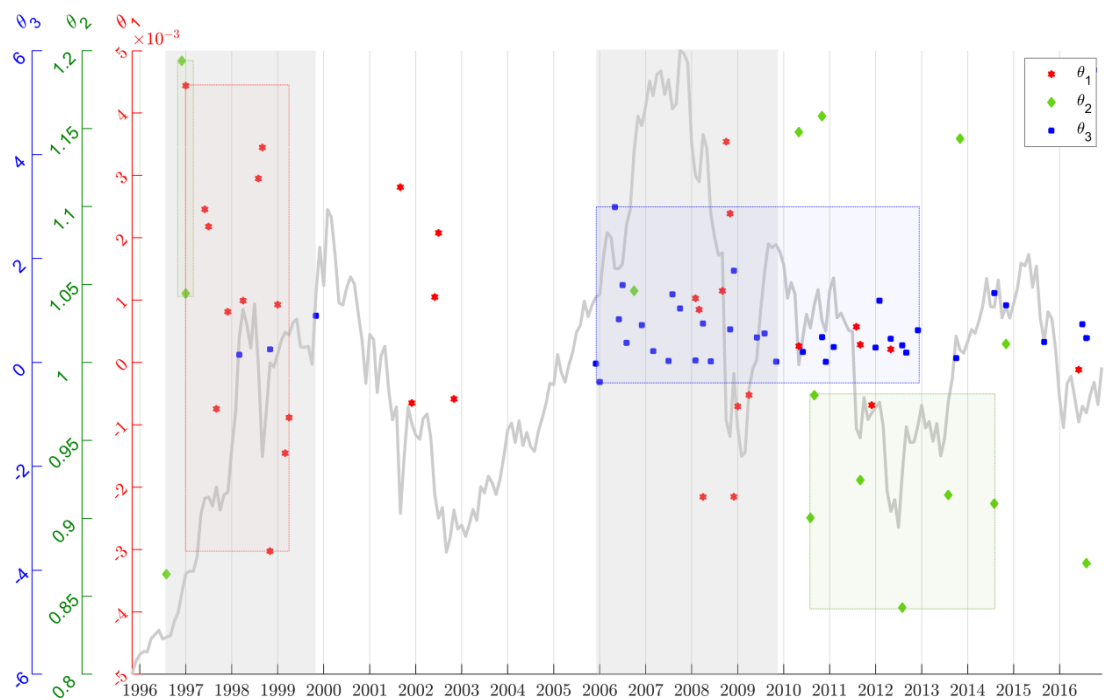

Notes: Shaded areas indicate a high concentration of sentiment corrections. Colored areas indicate a high proportion of investor optimism ($\theta_1$), overconfidence ($\theta_2$) or tail sentiment adjustments ($\theta_3$).

Three main observations can be drawn from Figure 2. First, our behavioral framework appropriately identifies the empirical phases that are generally associated with high sentiment periods. The interval from December 2005 to November 2009 contains 34% of our sentiment corrections and broadly coincides with the ramp-up and materialization of the 2007-08 global financial crisis (GFC). Likewise, the period from December 1996 to April 1999 comprises 19% of our adjustments and can be associated with the formation of the dot.com bubble. Both periods are broadly seen as phases where irrational behaviors may have led to pricing anomalies, hence validating our estimation framework.

---

[22] The values reported for $\theta_{1,t}$, $\theta_{2,t}$ and $\theta_{3,t}$ are derived from our low sentiment calibration. High sentiment figures can be obtained by linear transformation.

Second, the adjustments for each sentiment dimension are also concentrated around specific periods. The interval from June 1997 to April 1999 gathers over 34% of our investor optimism corrections. Behavioral adjustments over this period are driven by sharp increases in implied volatilities, indicating that despite the growing asset prices market participants were wary of the rise. However, the dot.com bubble is also characterized by several sharp decreases in implied volatilities, signaling investor confusion and rapid swings in investor sentiment during this period.

In terms of tail expectations, both left-tail fear and right-tail exuberance are observed from December 2005 to June 2008, delineating a high sentiment period in the lead up to the GFC. In contrast, once the GFC materialized, all tail adjustments stem from the excessive fear of substantial losses embedded in options markets (i.e. high negative skewness). Regarding investor confidence, although behavioral corrections are less concentrated over time, sharp decreases in trading volumes generate an intermittent phase of underconfidence from 2011 to 2014, whereas several increases in 1996 lead up to a short period of overconfidence.

Third, Figure 2 shows that our sentiment adjustments may change rapidly from positive to negative values. This result stems from our estimation framework; since our behavioral proxies are derived from forward-looking inputs that are updated daily, our adjustments track the day-to-day changes in investor sentiment, and can accommodate rapid changes in market expectations.

Table 2 shows the summary statistics of our sentiment proxies. For investor optimism and overconfidence, the 5$^{th}$ and 95$^{th}$ $t$-quantiles of the $\Delta IV_t$ and $\Delta TV_t$ time series mark the thresholds to activate our behavioral corrections. Therefore, monthly increases in implied volatilities higher than 10.6 points are linked to excessive pessimism, whereas falls beyond -9.4 points are indicative of excessive optimism. Similarly, increases in trading volumes higher than 51% are associated with overconfidence, whereas falls over -32% are linked to underconfidence[23]. Regarding tail sentiment, most $Skew_t$ values are concentrated around negative values, leading to higher corrections due to excessive left-tail fear compared with right-tail exuberance. The correlation among our sentiment proxies is remarkably low, indicating that our proxies contain specific information about complementary angles of investor sentiment.

Table 3 reports the number of corrections performed for each sentiment dimension. The adjustments stand at 32 for investor optimism, 15 for overconfidence and 39 for tail sentiment. However, as several corrections can be activated at the same time, we perform density transformations in 72 out of 254 monthly observations (28% of the sample).

---

[23] These figures should be understood as approximate values for the entire sample. Actual thresholds are time-variant and estimated *ex-ante* with the information available up to date $t$.

**Table 2: Summary statistics for the sentiment proxies**

| Sentiment dimension | Market proxy | Mean | Standard deviation | 5th percentile | 95th percentile |
|---|---|---|---|---|---|
| Investor Optimism | $\Delta IV_t$ | 0.000 | 0.072 | -0.094 | 0.106 |
| Investor Confidence | $\Delta TV_t$ | 1.036 | 0.252 | 0.676 | 1.515 |
| Tail Sentiment | $Skew_t$ | -0.976 | 0.771 | -2.227 | -0.082 |

| Correlation | $\Delta IV_t$ | $\Delta TV_t$ |
|---|---|---|
| $\Delta TV_t$ | 0.113 | |
| $Skew_t$ | 0.130 | 0.058 |

**Table 3: Behavioral corrections by sentiment dimension**

| Sentiment dimension | Adjustment | Monthly activations | Pessimism (optimism) | Underconfidence (overconfidence) | Left-tail fear (right-tail exuberance) |
|---|---|---|---|---|---|
| Investor Optimism | $\theta_{1,t}$ | 32 | 20 (12) | | |
| Investor Confidence | $\theta_{2,t}$ | 15 | | 8 (7) | |
| Tail Sentiment | $\theta_{3,t}$ | 39 | | | 37 (2) |

## 6.2 Integrated Forecast Score

We now examine the predictive ability of all forecast models. Forecast power is measured by the IFS, which aggregates in a standardized [0, 1] ranking the results from the log score, CRPS and goodness-of-fit analyses. For each density model, Figure 3 reports the IFS of our real-world forecasts compared with the IFS of the corresponding traditional variant.

Our results show that behavioral effects can be effectively used to forecast future prices. For all 15 underlying models, the best prediction is always obtained when sentiment is disentangled from fundamental expectations. Overall, 97% of our real-world forecasts outperform the corresponding no-sentiment model. The average IFS improvement is +0.05, which is 8 times higher than the decrease observed in the single case of no improvement.

Both our low and high sentiment calibrations deliver substantial forecast gains. Low sentiment adjustments generate information gains across all 15 traditional models, increasing the IFS by 0.057 on average. Similarly, our high sentiment correction improves 14 out of 15 traditional predictions, leading to an IFS gain of 0.043. Simpler stochastic models (i.e. BSM and Heston) perform relatively better with large behavioral corrections, whereas more sophisticated dynamics improve most with lower adjustments.

**Figure 3: IFS comparison: Real-world versus traditional densities**

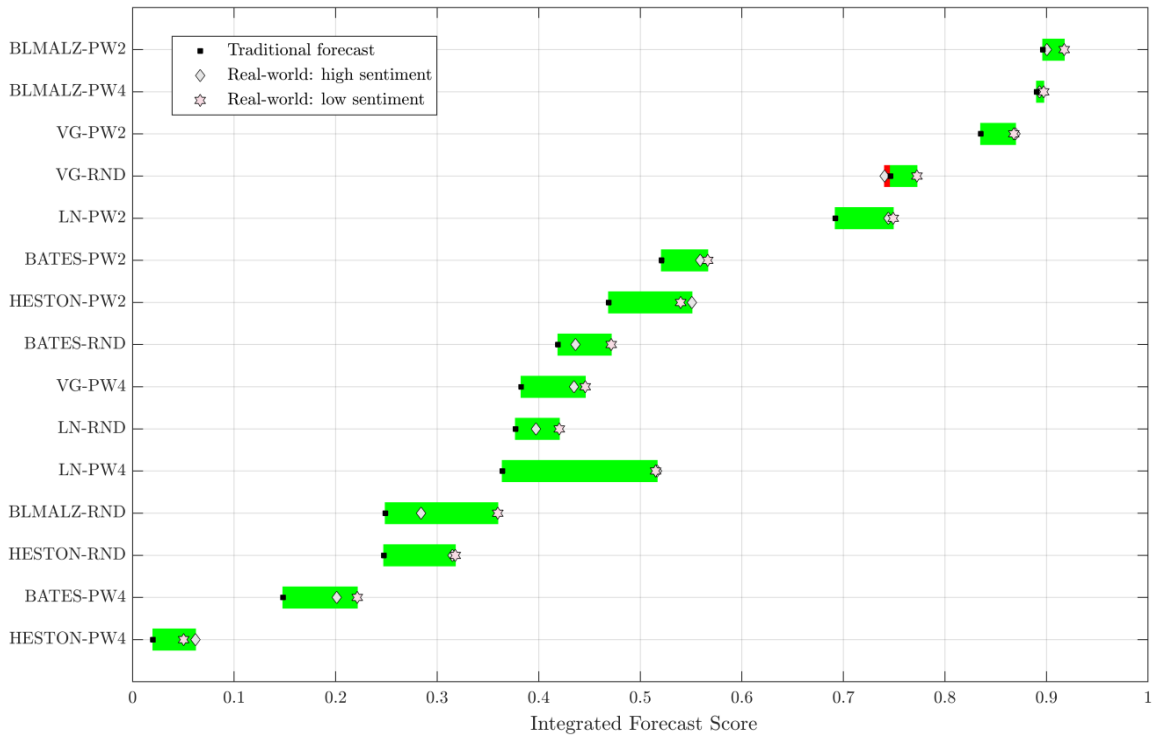

Notes: Horizontal bars represent alternative model specifications. Real-world adjustments which generate forecast gains are highlighted in green.

Before sentiment, densities estimated with alternative risk assumptions exhibit notably different performances; distributions with CRRA $\gamma = 2$ preferences deliver the highest overall IFS (0.68 on average), followed by RNDs (0.41) and CRRA $\gamma = 4$ densities (0.36). Remarkably, for all possible risk-assumptions, our behavioral correction significantly increases the predictive ability of the corresponding traditional models. The highest information gains are observed in the CRRA $\gamma = 4$ models (+0.063 incremental IFS), suggesting that sentiment effects can be particularly useful to correct possibly misspecified densities.

### 6.3 Log-likelihood comparisons

Table 4 summarizes the local accuracy achieved by our risk-neutral, risk-adjusted and real-world densities. Reported figures are the incremental log-likelihood of behavioral densities compared with the no-sentiment variants. Overall, the introduction of a sentiment correction improves the out-of-sample log-likelihood of traditional forecasts in 28 out of 30 of our real-world densities, generating an average increase of +1.05[24].

Forecast gains are robust across all risk-preferences, underlying models and sentiment calibrations. By risk preferences, log-likelihood gains stand at 0.87 for RNDs, 0.97 for CRRA $\gamma = 2$ and 1.31 for CRRA $\gamma = 4$ densities. Regarding sentiment effects, our low impact calibration outperforms the large correction (1.12 vs. 0.97). However, the highest accuracy gains are observed through the high sentiment correction of the CRRA $\gamma = 4$ densities (1.37), indicating again that behavioral effects can correct possibly misspecified risk-preferences.

---

[24] Note that loglikelihood gains come from behavioral adjustments in only 72 out of 254 monthly dates, whereas overall loglikelihood are derived from the entire sample figures.

Table 4: Out-of-sample log-likelihood

| Underlying model | Traditional forecast | Real-world densities | |
|---|---|---|---|
| | No Sentiment | Low Sentiment | High Sentiment |
| **Risk-neutral** | | | |
| LN-RND | -1815.19 | 0.76 | 0.44 |
| HESTON-RND | -1817.36 | 1.27 | 1.29 |
| BATES-RND | -1819.50 | 1.02 | 0.91 |
| VG-RND | -1812.85 | 0.85 | 0.51 |
| BLMALZ-RND | -1816.47 | 1.48 | 0.14 |
| **CRRA $\gamma = 2$** | | | |
| LN-PW2 | -1813.73 | 1.12 | 1.01 |
| HESTON-PW2 | -1816.21 | 1.36 | 1.61 |
| BATES-PW2 | -1819.13 | 1.14 | 1.29 |
| VG-PW2 | -1811.41 | 0.86 | 0.77 |
| BLMALZ-PW2 | -1810.54 | 0.74 | -0.22 |
| **CRRA $\gamma = 4$** | | | |
| LN-PW4 | -1817.13 | 1.75 | 1.81 |
| HESTON-PW4 | -1820.78 | 1.74 | 2.17 |
| BATES-PW4 | -1823.44 | 1.48 | 1.82 |
| VG-PW4 | -1814.51 | 1.05 | 1.09 |
| BLMALZ-PW4 | -1810.15 | 0.24 | -0.04 |

Notes: Log-likelihood figures for all real-world densities are computed as the value in excess of the corresponding traditional forecast.

Among traditional forecasts, the risk-adjusted BLMALZ densities exhibit the highest accuracy, hence requiring lower overall corrections. Remarkably, even for the most accurate traditional models, our low sentiment calibration delivers significant information gains, increasing out-of-sample log-likelihoods by 0.49 on average.

### 6.4 CRPS comparisons

Table 5 reports the forecasting errors of all predictive schemes. Our results show that a simple behavioral adjustment significantly improves the CRPS of traditional forecasts.

The error reduction is robust across all underlying models, risk preferences and sentiment calibrations. By underlying dynamics, the highest CRPS decrease is observed in the Heston model (-0.0136 on average) whereas the lowest improvement corresponds to the VG process (-0.0026). Regarding risk preferences, the CRPS reduction is highest for the CRRA $\gamma = 4$ densities (-0.0159) followed by RNDs (-0.0072) and CRRA $\gamma = 2$ densities (-0.0054). Our low sentiment calibration reduces global errors by -0.0111, whereas the high sentiment correction generates an average decrease of -0.0078.

**Table 5: Continuous Ranked Probability Score**

| Underlying model | Traditional forecast | Real-world densities | |
| --- | --- | --- | --- |
| | No Sentiment | Low Sentiment | High Sentiment |
| **Risk-neutral** | | | |
| LN-RND | 3.480 | **-0.0047** | **-0.0012** |
| HESTON-RND | 3.498 | **-0.0101** | **-0.0006** |
| BATES-RND | 3.455 | **-0.0100** | **-0.0036** |
| VG-RND | 3.452 | **-0.0001** | **0.0068** |
| BLMALZ-RND | 3.523 | **-0.0280** | **-0.0165** |
| **CRRA $\gamma = 2$** | | | |
| LN-PW2 | 3.435 | **-0.0101** | **-0.0093** |
| HESTON-PW2 | 3.473 | **-0.0049** | **-0.0052** |
| BATES-PW2 | 3.430 | **-0.0099** | **-0.0036** |
| VG-PW2 | 3.440 | **-0.0045** | **-0.0057** |
| BLMALZ-PW2 | 3.410 | **-0.0023** | **0.0018** |
| **CRRA $\gamma = 4$** | | | |
| LN-PW4 | 3.489 | **-0.0265** | **-0.0260** |
| HESTON-PW4 | 3.581 | **-0.0223** | **-0.0300** |
| BATES-PW4 | 3.503 | **-0.0245** | **-0.0171** |
| VG-PW4 | 3.494 | **-0.0084** | **-0.0038** |
| BLMALZ-PW4 | 3.421 | **-0.0008** | **0.0005** |

Notes: CRPS figures for all real-world densities are computed as the value in excess of the corresponding traditional forecast.

CRPS improvements are broadly in line with the log-likelihood gains. The similarity stems from the typically mount-shaped distributions that characterize equity models. In such distributions, *ex-post* realizations falling near the peak of the mountain simultaneously increase the log-likelihood and reduce the distance to other probability masses, improving both the log score and the CRPS. However, this relationship is particularly affected by higher moments. For instance, the Heston model achieves better log scores but worse CRPS than the Bates model, which can be explained by the more negatively skewed distributions that are typically observed in our Bates density forecasts.

### 6.5 Statistical consistency

Table 6 summarizes the statistical consistency of our traditional and real-world densities. Reported figures are the incremental *p*-values of behavioral densities compared to the no-sentiment variants. Our results show that real-world densities provide better forecasts of the distribution of ex-post realizations than both risk-neutral and risk-adjusted predictions.

For the Berkowitz LR3 test, the improvement rate is 77% with an average *p*-value increase of 0.011 across all underlying models. Similarly, over 80% and 83% of our real-world densities generate improvements in the JB and KS tests, increasing *p*-values by 0.034 and 0.021, respectively.

**Table 6: Goodness-of-fit tests**

| Model | Traditional forecast | | | Low Sentiment | | | High Sentiment | | |
|---|---|---|---|---|---|---|---|---|---|
| | LR3 | JB | KS | LR3 | JB | KS | LR3 | JB | KS |
| **Risk-neutral** | | | | | | | | | |
| LN-RND | 0.175 | 0.001 | 0.016 | -0.007 | 0.000 | 0.000 | -0.008 | 0.000 | 0.000 |
| HESTON-RND | 0.090 | 0.001 | 0.008 | 0.009 | 0.000 | 0.004 | 0.013 | 0.000 | 0.004 |
| BATES-RND | 0.172 | 0.002 | 0.148 | -0.025 | -0.001 | 0.049 | -0.024 | -0.001 | 0.044 |
| VG-RND | 0.200 | 0.622 | 0.096 | 0.006 | 0.057 | 0.000 | 0.002 | 0.058 | -0.026 |
| BLMALZ-RND | 0.002 | 0.533 | 0.008 | 0.000 | 0.112 | 0.002 | 0.000 | 0.110 | 0.002 |
| **CRRA $\gamma = 2$** | | | | | | | | | |
| LN-PW2 | 0.600 | 0.001 | 0.684 | 0.060 | 0.000 | -0.120 | 0.078 | 0.000 | -0.135 |
| HESTON-PW2 | 0.274 | 0.001 | 0.683 | 0.045 | 0.000 | 0.086 | 0.056 | 0.000 | 0.088 |
| BATES-PW2 | 0.229 | 0.002 | 0.987 | -0.025 | -0.001 | 0.006 | -0.013 | -0.001 | 0.000 |
| VG-PW2 | 0.506 | 0.631 | 0.291 | 0.006 | 0.065 | 0.077 | 0.008 | 0.065 | 0.077 |
| BLMALZ-PW2 | 0.158 | 0.469 | 0.492 | 0.036 | 0.056 | 0.131 | 0.048 | 0.085 | 0.147 |
| **CRRA $\gamma = 4$** | | | | | | | | | |
| LN-PW4 | 0.054 | 0.001 | 0.157 | 0.030 | 0.000 | 0.116 | 0.036 | 0.000 | 0.139 |
| HESTON-PW4 | 0.009 | 0.001 | 0.012 | 0.004 | 0.000 | 0.004 | 0.005 | 0.000 | 0.006 |
| BATES-PW4 | 0.007 | 0.002 | 0.104 | 0.000 | -0.001 | 0.030 | 0.001 | -0.001 | 0.041 |
| VG-PW4 | 0.037 | 0.646 | 0.009 | 0.005 | 0.053 | 0.001 | 0.005 | 0.071 | 0.004 |
| BLMALZ-PW4 | 0.256 | 0.477 | 0.489 | -0.019 | 0.138 | -0.076 | 0.002 | 0.163 | -0.076 |

Notes: The *p*-values for all real-world densities are computed as the value in excess of the corresponding traditional forecast.

The incremental consistency is robust across all model classifications. By risk category, the combination of a moderate risk aversion and high sentiment effects deliver the best overall performance. By underlying model, the most prominent gains are observed in the Heston model, where all goodness-of-fit tests exhibit statistical improvements. Conversely, the LR3 and JB *p*-values exhibit decreases in the Bates model. This can be explained by an *ex-post* realization in September 2001 which becomes more extreme through our sentiment correction, affecting the T-PIT consistency due to small sample considerations. In contrast, when statistical improvement is measured with the KS test, all our real-world densities increase the corresponding *p*-values in all Bates specifications.

Remarkably, our high sentiment calibration outperforms the low sentiment adjustment in our goodness-of-fit tests. This contrast with the better performance of the lower calibration in both log-likelihood and CRPS, demonstrating that different metrics may lead to diverging model choices, and highlighting the importance of considering both accuracy and consistency in density forecast evaluations.

### 6.6 Robustness analyses

This section performs two additional analyses. First, we compare the predictive ability of our sentiment adjustments against non-parametric forecasts that are recalibrated to avoid past mistakes. Second, we explore the performance of our behavioral transformations when applied to traditional densities where an implied risk-aversion is estimated from option prices.

## 6.6.1 Statistical recalibration

Historical biases observed in density predictions can be corrected through a recalibration of the current forecast in light of past mistakes. Following Fackler and King (1990), Shackleton et al. (2010) and De Vincent-Humphreys and Noss (2012), we calculate the empirical calibration function $\hat{c}_t$ from the difference between the $T\text{-}PIT_t$ sequence of each traditional density, and the $T\text{-}PIT_t$ of a perfectly specified model:

$$\hat{c}_t(z_t) = \hat{h}_t(z_t) / \phi(z_t) \tag{35}$$

where $z_t = \Phi^{-1}(T\text{-}PIT_t)$, $\hat{h}_t$ represents a normal KDE, and $\phi$ denotes the probability density function of a standard normal variable. For each date $t$, the recalibrated distribution $f^{RC}$ is estimated from the risk-adjusted density $f^{RA}$ as:

$$f^{RC}(x_T) = f^{RA}(x_T)\hat{c}_t(z_t) \tag{36}$$

By construction, the calibration function reflects all PIT biases observed in past predictions, hence correcting both risk preferences and sentiment-induced mistakes. Consequently, to appropriately compare the performance of our behavioral framework with statistical recalibrations, we recalibrate traditional densities on the dates where sentiment corrections are performed.

Table 7 shows that our real-world densities outperform densities recalibrated to avoid past mistakes, delivering an average IFS increase of 0.05 compared with -0.02 in statistical recalibrations. However, the comparison shows notable differences across risk-preference assumptions. When the traditional forecast is risk-neutral, statistical recalibrations deliver higher information gains than our sentiment corrections. This can be attributed to the type of biases that are corrected through recalibrations; as RNDs lack both risk-preferences and sentiment effects, simultaneously correcting both generates larger IFS gains than just correcting behavioral mistakes.

Conversely, when the initial forecast is already risk-adjusted, our sentiment framework delivers higher gains than statistical recalibrations, demonstrating that forward-looking behavioral adjustments perform better than corrections based on past mistakes. Further analyses confirm that while statistical tests generally improve [25], PIT-based recalibrations do not generate log-likelihood or CRPS gains (-0.89 log score; +0.014 CRPS), reinforcing the use of composite measures that summarize both accuracy and statistical consistency.

---

[25] The average *p*-value increases are 0.18, 0.04 and 0.12 for the LR3, JB and KS tests, respectively.

**Table 7: IFS comparison: Recalibrated versus real-world densities**

| Underlying model | Traditional forecast | Recalibrated density | Real-world densities | |
|---|---|---|---|---|
| | | | Low Sentiment | High Sentiment |
| **Risk-neutral** | | | | |
| LN-RND | 0.377 | **0.14** | **0.04** | **0.02** |
| HESTON-RND | 0.247 | **0.13** | **0.07** | **0.07** |
| BATES-RND | 0.419 | **-0.03** | **0.05** | **0.02** |
| VG-RND | 0.746 | **-0.14** | **0.03** | **-0.01** |
| BLMALZ-RND | 0.249 | **0.34** | **0.11** | **0.05** |
| **CRRA $\gamma = 2$** | | | | |
| LN-PW2 | 0.692 | **-0.06** | **0.06** | **0.05** |
| HESTON-PW2 | 0.469 | **-0.06** | **0.07** | **0.08** |
| BATES-PW2 | 0.521 | **-0.12** | **0.05** | **0.04** |
| VG-PW2 | 0.835 | **-0.17** | **0.03** | **0.03** |
| BLMALZ-PW2 | 0.896 | **-0.08** | **0.02** | **0.00** |
| **CRRA $\gamma = 4$** | | | | |
| LN-PW4 | 0.364 | **-0.01** | **0.15** | **0.15** |
| HESTON-PW4 | 0.020 | **0.01** | **0.03** | **0.04** |
| BATES-PW4 | 0.148 | **-0.03** | **0.07** | **0.05** |
| VG-PW4 | 0.383 | **-0.05** | **0.06** | **0.05** |
| BLMALZ-PW4 | 0.890 | **-0.19** | **0.01** | **0.00** |

Notes: IFS figures for all recalibrated and real-world densities are computed as the value in excess of the corresponding traditional forecast.

### 6.6.2 Densities with option-implied risk preferences

Following Bakshi and Madan (2006) and Kang et al. (2010), the option-implied relative risk aversion (IRRA) at date $t$ can be obtained from the difference between the risk-neutral and the physical variance. Specifically, when the representative investor features a power utility function, the IRRA coefficient $\gamma_t$ can be extracted from the equation:

$$\frac{\sigma_{p,t}^2(\tau) - \sigma_{q,t}^2(\tau)}{\sigma_{q,t}^2(\tau)} \approx \gamma_t \sigma_{q,t}(\tau) \xi_{q,t}(\tau) + \frac{\gamma_t^2}{2} \sigma_{q,t}^2(\tau)(\kappa_{q,t}(\tau) - 3) \qquad (37)$$

where $\sigma_{q,t}^2$, $\xi_{q,t}$, and $\kappa_{q,t}$ are the variance, skewness and kurtosis of the risk-neutral distribution, and $\sigma_{p,t}^2(\tau)$ is the physical variance computed from the last 30 trading days. All risk-neutral moments are obtained using Bakshi et al. (2003) formulae.

Instead of using historical moments, Equation (38) employs the forward-looking skewness and kurtosis, producing IRRA estimates that are consistent with our option-implied densities. For each date $t$, we obtain $\gamma_t$ by minimizing the objective function:

$$\varepsilon_t = \frac{\sigma_{p,t}^2(\tau) - \sigma_{q,t}^2(\tau)}{\sigma_{q,t}^2(\tau)} - \gamma_t \sigma_{q,t}(\tau) \xi_{q,t}(\tau) - \frac{\gamma_t^2}{2} \sigma_{q,t}^2(\tau)(\kappa_{q,t}(\tau) - 3) \qquad (38)$$

Given the challenges in obtaining reliable IRRA estimates, we restrict the range of acceptable $\gamma_t$ values from -1 to 6. This choice generates additional flexibility compared to our initial $\gamma$ range (0, 2 and 4), while avoiding extreme swings and economically implausible IRRA values.

Table 8 shows that our real-world densities consistently improve the predictive power of forecasts obtained with IRRA. When real-world densities are estimated with low sentiment effects, our behavioral correction improves the IFS of all traditional models, delivering an average IFS gain of 0.04. Similarly, when high sentiment effects are employed, 4 out of 5 behavioral corrections improve the corresponding no-sentiment variants.

**Table 8: IFS comparison: Densities with option-implied risk preferences**

| Underlying model | Traditional forecast | Real-world densities | |
|---|---|---|---|
| | No sentiment | Low Sentiment | High Sentiment |
| **Time-variant IRRA** | | | |
| LN-IRRA | 0.56 | **0.11** | **0.07** |
| HESTON-IRRA | 0.29 | **0.08** | **0.06** |
| BATES-IRRA | 0.43 | **0.03** | **0.02** |
| VG-IRRA | 0.78 | **0.00** | **-0.03** |
| BLMALZ-IRRA | 0.82 | **0.06** | **0.01** |

Notes: IFS figures for real-world densities are computed as the value in excess of the corresponding traditional forecast.

Information gains are driven by improvements in both accuracy and statistical consistency. The incremental log-likelihood stands at +0.94, the CRPS reduction is -0.005 and the average *p*-value increase across all statistical tests is +0.03, demonstrating that our real-world forecasts also improve neoclassical models where a time-variant risk aversion is obtained from option prices.

# 7. Conclusion

This paper examines whether investor sentiment can be used to improve the forecasting ability of density predictions obtained from option prices. Increasing evidence shows that real-world investors commit systematic behavioral errors that manifest in asset prices. Consequently, it follows that market-implied forecasts should be appropriately corrected to disentangle the impact of behavioral biases from fundamental expectations.

To quantify sentiment effects, we develop a forward-looking framework that generates the behavioral correction required to adjust traditional forecasts in specific areas of the return distribution. For 15 underlying models and risk-preference combinations, we show that a simple behavioral transformation in the mean, variance and tail estimates of traditional predictions significantly improve their accuracy and statistical consistency.

Information gains are robust across all forecast metrics and sentiment calibrations, demonstrating that behavioral effects can be effectively used to predict asset prices. Our results also show that real-world densities outperform non-parametric corrections derived from past mistakes, and improve forecast models where risk aversion is dynamically estimated from option prices.